\documentclass[aps,prl,twocolumn,showpacs,floatfix,superscriptaddress,nofootinbib,nobibnotes,longbibliography]{revtex4-1}
\usepackage[utf8]{inputenc}
\usepackage{amsmath}
\usepackage{amssymb}
\usepackage{mathtools}
\usepackage{graphicx}
\usepackage{hyperref}
\usepackage{color}
\usepackage{multirow}

\graphicspath{{.}}

\begin{document}

\title{Corrections to nucleon capture cross sections computed in
  truncated Hilbert spaces}
  
\author{B.~Acharya}
\affiliation{Department of Physics and Astronomy,
University of Tennessee, Knoxville, TN 37996, USA}

\author{A.~Ekstr\"om}
\affiliation{Department of Physics, Chalmers University of Technology, SE-412 96 G\"oteborg, Sweden}

\author{D.~Odell}
\affiliation{Department of Physics and Astronomy,
University of Tennessee, Knoxville, TN 37996, USA}

\author{T.~Papenbrock}
\affiliation{Department of Physics and Astronomy,
University of Tennessee, Knoxville, TN 37996, USA}
\affiliation{Physics Division, Oak Ridge National Laboratory, Oak Ridge, TN 37831, USA}\thanks{This manuscript has been authored
  by UT-Battelle, LLC under Contract No. DE-AC05-00OR22725 with the
  U.S. Department of Energy. The United States Government retains and
  the publisher, by accepting the article for publication,
  acknowledges that the United States Government retains a
  non-exclusive, paid-up, irrevocable, world-wide license to publish
  or reproduce the published form of this manuscript, or allow others
  to do so, for United States Government purposes. The Department of
  Energy will provide public access to these results of federally
  sponsored research in accordance with the DOE Public Access
  Plan. (http://energy.gov/downloads/doe-public-access-plan).}

\author{L.~Platter}
\affiliation{Department of Physics and Astronomy,
University of Tennessee, Knoxville, TN 37996, USA}
\affiliation{Physics Division, Oak Ridge National Laboratory, Oak Ridge, TN 37831, USA}

\date{\today}
\begin{abstract}
  Nucleon capture cross sections enter various astrophysical
  processes. The measurement of proton capture on nuclei at
  astrophysically relevant low energies is a challenge, 
and theoretical computations in this long-wavelength regime are sensitive to 
  the long-distance asymptotics of the wave functions. A theoretical foundation 
  for estimating and correcting errors introduced in capture cross sections 
  due to Hilbert space truncation has so far been lacking. 
  We derive extrapolation formulas that relate the infrared
  regularized capture amplitudes to the infinite basis limit and
  demonstrate their efficacy for proton-proton fusion.  Our results
  are thus relevant to current calculations of few-body capture
  reactions such as proton-proton fusion or proton capture on the
  deuteron, and they also open the way for 
the use of {\it ab initio} 
many-body wave functions represented in finite Hilbert spaces in precision calculations of 
  nucleon capture on heavier nuclei.
 
\end{abstract}

\maketitle

\paragraph{Introduction.} Processes in which a nucleon is captured by
a nucleus occur in many areas of pure and applied physics. They play
an important role in big bang nucleosynthesis and in the nuclear
astrophysics of stars, novae, X-ray bursts and supernovae, see,
e.g. Ref.~\cite{bertulani2016} for a recent review.  Capture reaction
rates are essential inputs for computations of stellar
models~\cite{adelberger2011}.  Proton capture cross sections are very
difficult to measure at astrophysically relevant energies below the
Coulomb barrier, forcing us to rely on theoretical results. Here, {\it
  ab initio}
computations~\cite{girlanda2010,navratil2011,quaglioni2012,marcucci2013}
and studies based on effective field
theory~\cite{zhang2014,Ryberg:2014exa,Ryberg:2015lea,zhang2015,acharya2016}
aim at achieving model-independent results with reliable uncertainty
estimates.

We note that precise theoretical calculations are a challenge too,
because the regime of low-energies and long de Broglie wave lengths
requires one to employ very large Hilbert spaces. It is therefore
important to control the uncertainties in theoretical calculations of
cross sections that are due to limitations of finite model
spaces. This is the purpose of this work.  Let us
consider proton-proton fusion, i.e., $p + p \rightarrow d +\nu_e
+e^{+}$, as the simplest example of a proton capture reaction.  This
reaction has been studied extensively and a calculation that reduces
the uncertainty well below 1~\% would be an important
development~\cite{adelberger2011,marcucci2013,acharya2016}. As we will
see below, the corrections due to finite Hilbert spaces become
relevant if such a precision is aimed at in {\it ab initio}
computations.

Truncation of the Hilbert space imposes ultraviolet (UV) and infrared (IR) momentum
cutoffs~\cite{stetcu2007,hagen2010b,jurgenson2011,coon2012}, leading
to systematic errors in observables. Thus, capture reactions into
bound states computed in finite Hilbert spaces will suffer from 
truncation errors regardless of how well the continuum is treated.
An example of previous corrections of such shortcomings is
presented in Ref.~\cite{navratil2006}. Formulas for extrapolation of various
bound-state observables to the infinite-basis limit were derived in
Refs. \cite{furnstahl2012,more2013,koenig2014,odell2016}.  In the same
spirit, we study and quantify the IR corrections to the capture and
fusion cross sections calculated from wave functions represented in
truncated Hilbert spaces. We make use of the dependence of the IR
length scale $L$ on the parameters of the oscillator basis, which is
known for the two-body problem~\cite{more2013}, the no-core shell
model~\cite{wendt2015}, and many-body product
spaces~\cite{furnstahl2015}. Below, we also present the IR length
relevant for hyperspherical harmonics with Laguerre polynomials as
radial wave functions.

Recent progress in {\it ab initio} computations of
  reactions and scattering
  states~\cite{nollett2007,quaglioni2008,hagen2012c,fossez2015,dohet2016,shirokov2016,calci2016}
  (see also Ref.~\cite{navratil2016} for a recent review) has made it
  possible to calculate capture cross sections of medium mass and
  heavy nuclei using discrete-basis representations of bound state
  wave functions. This makes it a timely issue to understand and
  correct the shortcomings pertaining to the finite Hilbert space
  treatment of the bound states involved. 

\paragraph{Theoretical derivation.} In what follows, we focus on 
the nucleon-nucleon (NN) processes as examples where the truncation
error can be fully understood.  This allows us to derive IR
extrapolation formulas that have a more general applicability.  The
generalization to heavier nuclei will be discussed below.  We assume
that the nuclear interaction vanishes beyond the range $R$. Thus, at
relative distances $r\ge R$ the bound state radial wave function
calculated in a truncated basis has the asymptotic
form~\cite{more2013}
\begin{equation}
\label{eq:boundwfn}
u^{(L)}(r) \rightarrow A_\infty e^{-\gamma_\infty r} \left[1-e^{-2\gamma_\infty (L-r)}\right]~. 
\end{equation}
Here, $\gamma_\infty$ and $A_\infty$ are, respectively, the binding
momentum and the asymptotic normalization coefficient in the infinite
volume limit~\cite{more2013}. Equation~\eqref{eq:boundwfn} is
asymptotically valid for all partial waves.  However, its higher order
corrections for $s$-wave are of $\mathcal{O}(e^{-\gamma_\infty
  (2L+r)})$, much smaller than the $\mathcal{O}[1/(\gamma_\infty r)]$
corrections for higher partial waves.

Calculations of capture cross sections in a truncated basis, 
therefore, effectively involve the radial matrix elements 
\begin{equation}
\label{eq:radialintegral}
\mathcal{I}_\lambda (k;\eta;L) \equiv \int_0^L \mathrm{d}r\, u^{(L)}(r)\,r^\lambda\,u_k(r)~,
\end{equation} 
where $k$ is the momentum of the scattering wave function $u_k(r)$ in
the initial state. $\eta$ is the Sommerfeld parameter and $\lambda$ defines the 
multipolarity of the transition.  For an
electromagnetic capture process, the multipolarity is equal to
$\lambda$ for electric transitions and to $\lambda+1$ for magnetic transitions. 
For the weak process, the dominant contribution at low
energies comes from $\mathcal{I}_0 (k;\eta;L)$.

At $r \ge R$ and $kr\gg\eta$, the radial wave function of the initial state has the form
\begin{align}
\label{eq:scatteringwfn}
u_k(r) \rightarrow &\cos\delta_l \, \sin \left[kr-\eta\log(2kr)+\sigma_l-\frac{\pi l}{2}\right] \nonumber\\
&+ \sin\delta_l \, \cos\left[kr-\eta\log(2kr)+\sigma_l-\frac{\pi l}{2}\right]~,
\end{align}
with $\sigma_l$ being the Coulomb phase shift. For the case of neutron
capture, $\sigma_l = 0 = \eta$.  Apart from the subleading $\eta$
dependence, Eq.~\eqref{eq:scatteringwfn} has additional
$\mathcal{O}(1/(kr))$ corrections for $l>0$ even in the absence of
Coulomb interaction.

We now proceed to derive the IR truncation error,
  $\Delta\mathcal{I}_\lambda (k;\eta;L)$, in the matrix element
  $\mathcal{I}_\lambda (k;\eta;L)$ calculated in Hilbert spaces with
  $L\gg R$. However, in order to use the asymptotically valid
  approximations for the wave functions given in
  Eqs.~\eqref{eq:boundwfn} and \eqref{eq:scatteringwfn}, we
  additionally require $kL \gg \eta$ for proton capture and fusion
  reactions, and $kL \gg l$ for capture in partial waves with orbital
  angular momentum $l$.

We begin by splitting the radial integral, Eq.~\eqref{eq:radialintegral}, into two regions,
\begin{equation}
 \label{eq:split}
  \mathcal{I}_\lambda (k;\eta;L) = \left( \int_0^{R} + \int_{R}^L \right) \mathrm{d}r\, u_L(r)\,r^\lambda\,u_k(r)~.
\end{equation}
The second integral, which is independent of the details of the
nuclear interaction, can be evaluated analytically using
Eqs.~\eqref{eq:boundwfn} and \eqref{eq:scatteringwfn} to give
\begin{align}
\label{eq:second}
  \int_{R}^L \mathrm{d}r\, u^{(L)}(r)\, r^\lambda\,u_k(r) = &\int_{R}^\infty \mathrm{d}r\, u^{(\infty)}(r)\,r^\lambda\,u_k(r) \nonumber\\
  & + 2 \, {\rm Re}\left[ f_\lambda (k;\eta;L)\right]~,
\end{align}
where $u^{(\infty)}(r)$ is $u^{(L)}(r)$ at $L\rightarrow\infty$, and
\begin{align}
\label{eq:fformula}
f_\lambda (k;\eta;L)  = \frac{i}{2} & A_\infty\, e^{i(\delta_l+\sigma_l-\pi l/2)} \, (2k)^{-i\eta} \nonumber\\
&[ (\gamma_\infty-ik)^{(-\lambda-1+i\eta)}\nonumber\\
&\quad \times \Gamma (\lambda+1-i\eta,\gamma_\infty L-ikL) \nonumber\\
&~-e^{-2\gamma_\infty L} (-\gamma_\infty-ik)^{(-\lambda-1+i\eta)}\nonumber\\ 
&\quad \times \Gamma (\lambda+1-i\eta,-\gamma_\infty L-ikL) ]~,
\end{align}
is the result of an overlap integral of the asymptotic incoming and
outgoing scattering wave function with the finite volume bound state
wave function.

Here $\Gamma(c,z)$ is the complex continuation of the incomplete Gamma
function~\cite{gradshteyn}. We have dropped terms of
$\mathcal{O}(e^{-\gamma_\infty(2L-R)})$ in Eq.~\eqref{eq:second}. For
asymptotically large values of $\gamma_\infty L$, we can also replace
$u^{(L)}(r)$ in the first integral in Eq.~\eqref{eq:split}, which includes
the contribution from the $r\ll L$ region, by $u^{(\infty)}(r)$.
Equation~\eqref{eq:split} can then be written as
\begin{align}
 \label{eq:general}
 \Delta\mathcal{I}_\lambda (k;\eta;L) &=  \mathcal{I}_\lambda (k;\eta;\infty) - \, \mathcal{I}_\lambda (k;\eta;L) \nonumber\\
 &= - 2 \, {\rm Re}\left[ f_\lambda (k;\eta;L)\right]~,
\end{align}
where 
\begin{equation}
   \mathcal{I}_\lambda (k;\eta;\infty) = \int_0^\infty\mathrm{d}r\, u^{(\infty)}(r)\,r^\lambda\,u_k(r)
\end{equation}
is the radial matrix element $\mathcal{I}_\lambda (k;\eta;L)$ at $L\rightarrow\infty$.

In addition to the exponentially suppressed term we explicitly dropped
above, we have also neglected the contributions to
$\Delta\mathcal{I}_\lambda(k;\eta;L)$ from the higher-order $\eta$
dependence and higher partial wave corrections to
Eqs.~\eqref{eq:boundwfn} and \eqref{eq:scatteringwfn}.  These terms
scale as $\Delta\mathcal{I}_{\lambda-1}(k;\eta;L)$ and are therefore
only suppressed by a factor of $1/L$.  Using the leading order
approximation in the asymptotic expansion of the $\Gamma$ function,
\begin{equation}
\label{eq:gamm_exp}
\Gamma(c,z) = z^{c-1}\,e^{-z}\left(1+\frac{c-1}{z}+\ldots\right)~,
\end{equation}
valid for $|z| \gg 1$ and $|\arg z| < 3\pi/2$, in
Eq.~\eqref{eq:fformula}, the IR truncation error in the capture matrix
element reduces to a much simpler form,
\begin{align}
  \label{eq:asympt}
  \MoveEqLeft \Delta\mathcal{I}_\lambda (k;\eta;L) = \frac{2A_\infty \gamma_\infty}{\gamma_\infty^2+k^2}
  L^\lambda e^{-\gamma_\infty L} \nonumber\\
  & \times \sin\left(\delta_l+\sigma_l-{\pi l\over 2} +kL -\eta\log{2kL}\right)~, 
\end{align}
for asymptotically large values of $\gamma_\infty L$. We note that the
approximation for $\Gamma(c,z)$ used here in order to arrive at
Eq.~\eqref{eq:asympt} is exact for $\lambda=0$ neutron capture.
However, at larger values of $\lambda$ and $\eta$, this approximation
gets worse and it may be necessary to obtain the IR truncation error
using Eqs.~\eqref{eq:fformula} and \eqref{eq:general} instead.

Since the relative error in the cross section is twice that in the
matrix element, we find from Eq.~\eqref{eq:asympt} that the IR
truncation error in the cross section scales as $e^{-\gamma_\infty
  L}$. We note that this $e^{-\gamma_\infty L}$ convergence with
increasing $\gamma_\infty L$ is much slower than the
$e^{-2\gamma_\infty L}$ behavior found for bound-state observables
such as energies and radii~\cite{furnstahl2012}.

The extrapolation formula, Eq.~\eqref{eq:general}, and its asymptotic
form, Eq.~\eqref{eq:asympt}, are the main results of this work.  These
equations hold for heavier nuclei and for all reasonable models of the
nuclear Hamiltonian because the single-particle wave functions have
the asymptotic forms given in Eqs.~\eqref{eq:boundwfn} and
\eqref{eq:scatteringwfn} in the range $R\le r<L$.  They are valid for
neutron capture as well as for proton capture unless the energy is low
enough to warrant the use of Coulomb wave functions $F_l(kr)$ and
$G_l(kr)$ for all $r\lesssim L$ instead of the sine and the cosine
functions in Eq.~\eqref{eq:scatteringwfn}.  Furthermore, the same
radial matrix elements contribute to break-up cross sections as well.

\paragraph{Numerical results.}

For numerical calculations, we use the chiral effective field theory
(EFT) interaction from Ref.~\cite{entem2003}. We obtain the $pp$ and
$np$ scattering states by solving the momentum-space Schr\"odinger
equation.  We then calculate the radial matrix elements,
$\mathcal{I}_\lambda(k;\eta;L)$, numerically for a range of $L$ values
by expanding the deuteron wave function in HO bases of varying
dimensionality.

 In the IR regime, the finite harmonic oscillator basis we use is indistinguishable
  from a spherical box with radius~\cite{more2013}
\begin{equation}
  \label{eq:Lfull}
  L = \sqrt{2(N + 3/2 + 2)}\,b~.
\end{equation}
Here $N$ is the maximum number of oscillator quanta and
$b=\sqrt{1/(\mu \Omega)}$, the oscillator length for a system with
reduced mass $\mu$ and oscillator frequency $\Omega$, respectively. 
The hyperspherical harmonics basis is popular in few-body
problems~\cite{avery1989,leidemann2013} and has also been used in the
computation of capture reactions~\cite{girlanda2010,marcucci2013}. For
this reason, we also discuss the IR length $L_{\rm HH}$ relevant for
this method. Using the hyperradius $\rho$ and a momentum scale
$\beta$, the hyperradial basis functions \begin{equation*}
  \sqrt{m!\beta^{3A-3}\over\Gamma(a+m+1)} (\beta\rho)^{a-3A+4\over 2}
  e^{-{1\over 2}\beta\rho}L_m^a(\beta\rho)
\end{equation*}
are orthonormal under the hyperradial integration measure
$d\rho \rho^{3A-4}$, which is adequate for a translationally invariant
$A$-body system~\cite{barnea1999,bacca2012}. Here $L_m^a$ denotes the
associated Laguerre polynomial and $a$ is a parameter. Noting the
similarity between the hyperradial wave functions and the radial wave
functions of the three-dimensional harmonic oscillator,
i.e. identifying $a=l+1/2$ and $N=2n+l$ in Eq.~(\ref{eq:Lfull}), 
where $n$ is the largest degree of the Laguerre polynomial used, 
we infer that the IR length for the hyperspherical radial basis is
\begin{equation}
 L_{\rm HH} = (4n+2a+6)\beta^{-1} \ .
\end{equation}

For the NN processes, it is computationally feasible to calculate
$\mathcal{I}_\lambda(k;\eta;L)$ in a large enough basis and obtain an
accurate numerical approximation to
$\mathcal{I}_\lambda(k;\eta;\infty)$.  We begin by comparing the
numerical truncation error, $\Delta\mathcal{I}_\lambda (k;\eta;L)$,
thus obtained with those predicted by Eq.~\eqref{eq:asympt}.

\begin{figure}[t]
    \centering
    \includegraphics[width=0.98\columnwidth]{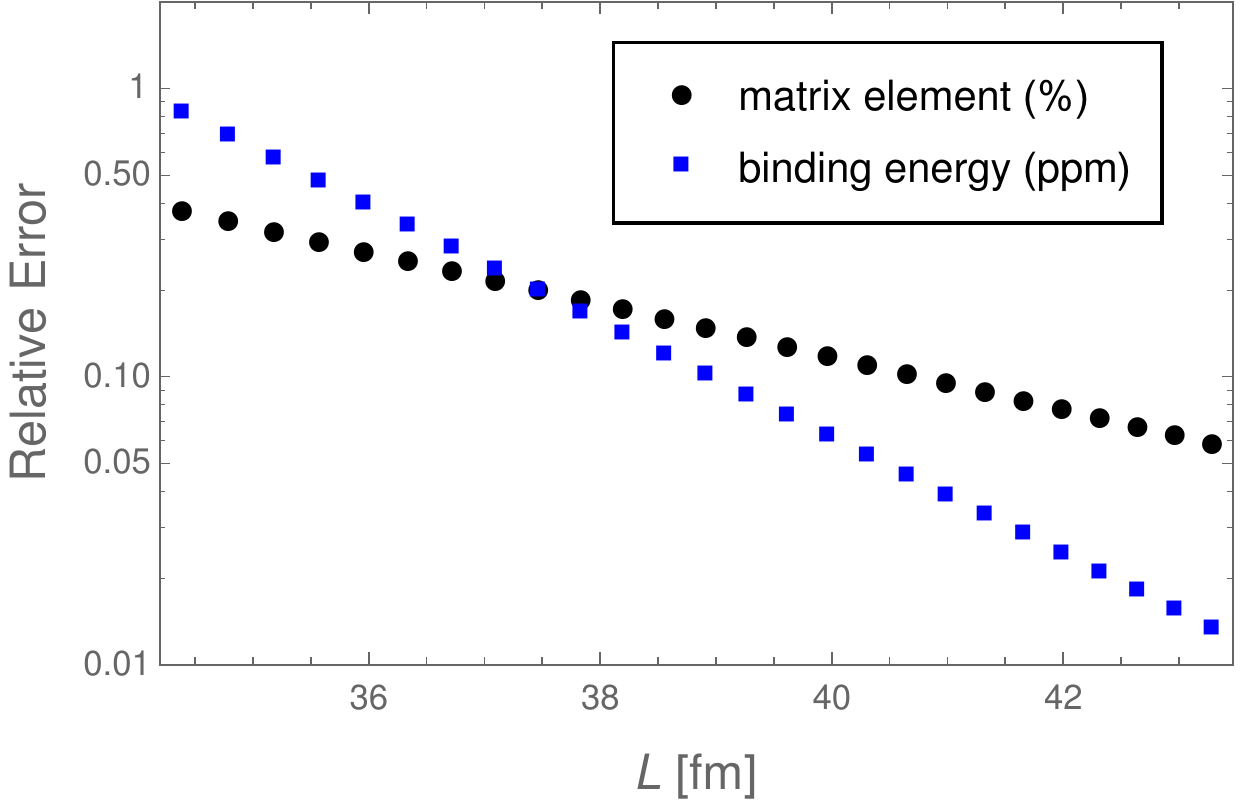}
    \caption{\label{fig:pp_error} IR truncation error in the
      axial-vector matrix element of $pp$ fusion at 50~keV
      center-of-mass energy (black circles) and in the deuteron
      binding energy (blue squares). The relative errors for the
      axial-vector matrix elements and those for the binding energy
      are expressed in percentage and parts per million (ppm)
      respectively.}
\end{figure}

In Fig.~\ref{fig:pp_error}, we plot the relative error due to a IR cutoff
in the matrix element of the Gamow-Teller operator between the deuteron $s$-wave
and the $pp$ $^1S_0$ wave functions at 50~keV center-of-mass energy,
$\Delta\mathcal{I}_0(k;\eta;L)/\mathcal{I}_0(k;\eta;\infty)$.  For
comparison, we also show the relative IR truncation error in the
deuteron binding energy.  The error in the matrix element at $L=35$~fm
is about 0.3~\%, which translates to an error in the cross section of about 0.6~\%. 
The size of this error is relevant for computing
$pp$ fusion cross sections to percentage precision, which recent calculations~\cite{marcucci2013}
aim at. In contrast, the deuteron binding
energy shows a much faster IR convergence --- the relative error at $L=35$~fm is 
about $0.5\times 10^{-6}$ --- reinforcing our claim that a basis that gives highly accurate results for bound state observables may 
still yield large systematic errors in capture cross section calculations. 
Furthermore, we have checked and verified that the
$L$-dependences of these errors are consistent with theoretical
predictions --- approximate $e^{-\gamma_\infty L}$ behavior for the
capture matrix element as derived above, and $e^{-2 \gamma_\infty L}$
for deuteron binding energy~\cite{more2013}.

\begin{figure}[t]
     \centering
     \includegraphics[width=0.98\columnwidth]{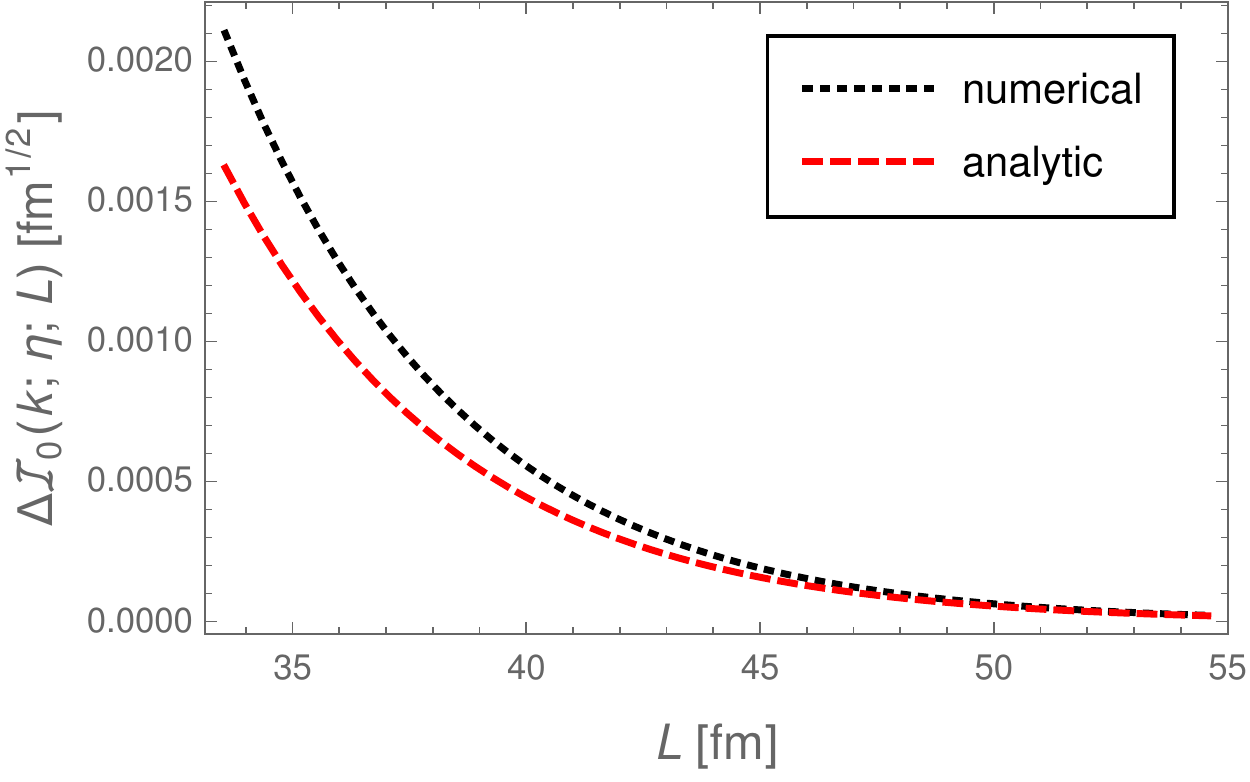}
     \caption{\label{fig:pp100k} Numerical and analytic results for the correction  
       $\Delta\mathcal{I}_0(k;\eta;L)$ to the radial overlap between the $^1S_0$ $pp$
       scattering wave function at 50~keV center-of-mass energy and the
       deuteron $s$-wave state.}
 \end{figure}

In Fig.~\ref{fig:pp100k}, we plot the truncation error for the $pp$
fusion matrix element shown earlier in Fig.~\ref{fig:pp_error} along
with the analytic result given by Eq.~\eqref{eq:asympt}.  Since
$\eta=0.5$ is not particularly small, we get a good agreement between
the analytic formula and numerical data at larger values of $L$, where
the corrections to Eq.~\eqref{eq:asympt} due to higher-order
$\eta$-dependence are less important.

\begin{figure}[t]
     \centering
     \includegraphics[width=0.98\columnwidth]{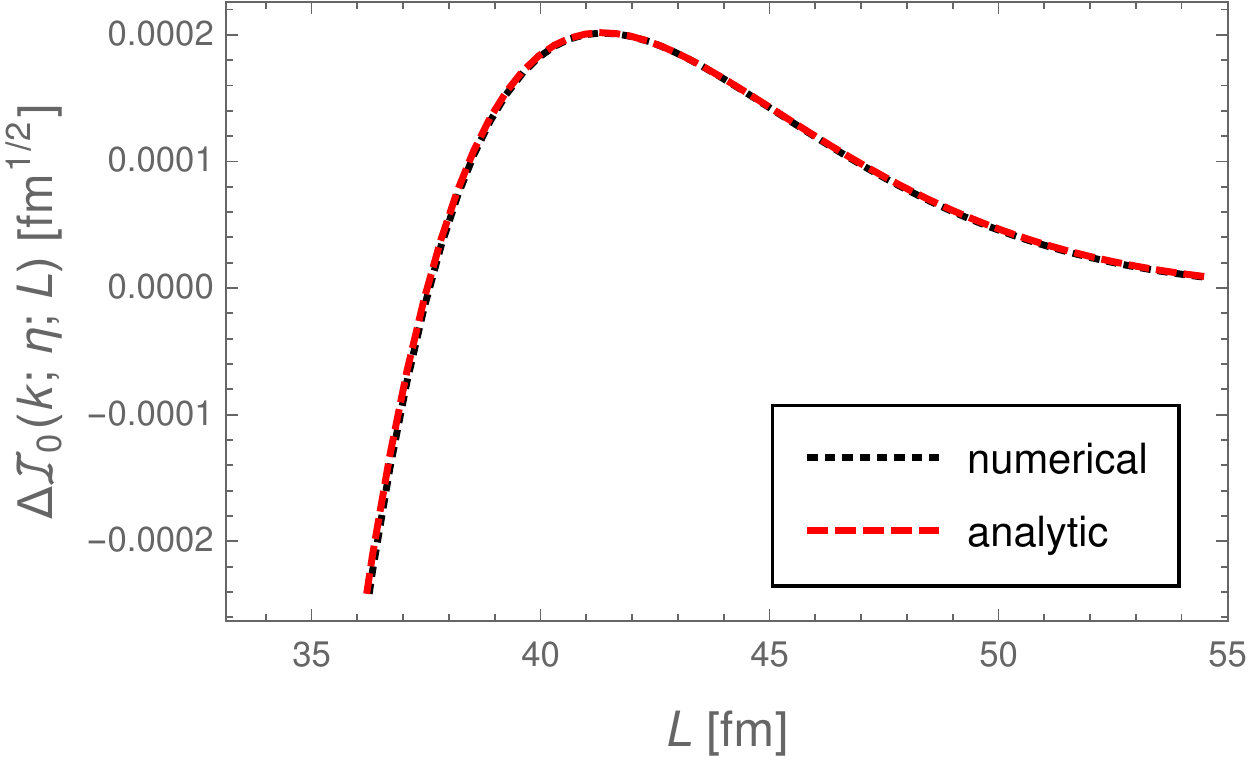}
     \caption{\label{fig:pp2M} Numerical and analytic results for the correction 
       $\Delta\mathcal{I}_0(k;\eta;L)$ to the radial overlap between the $^1S_0$ $pp$
       scattering wave function at 1~MeV center-of-mass energy and the
       deuteron $s$-wave state.}
 \end{figure}

For comparison, we plot $\Delta\mathcal{I}_\lambda(k;\eta;L)$ for
the same process at 1~MeV center-of-mass energy for the same range of
$L$ values in Fig.~\ref{fig:pp2M}. Since $\eta=0.11\ll1$ at this
energy, we find a much better agreement even at smaller $L$.

\begin{figure}[t]
     \centering
     \includegraphics[width=0.98\columnwidth]{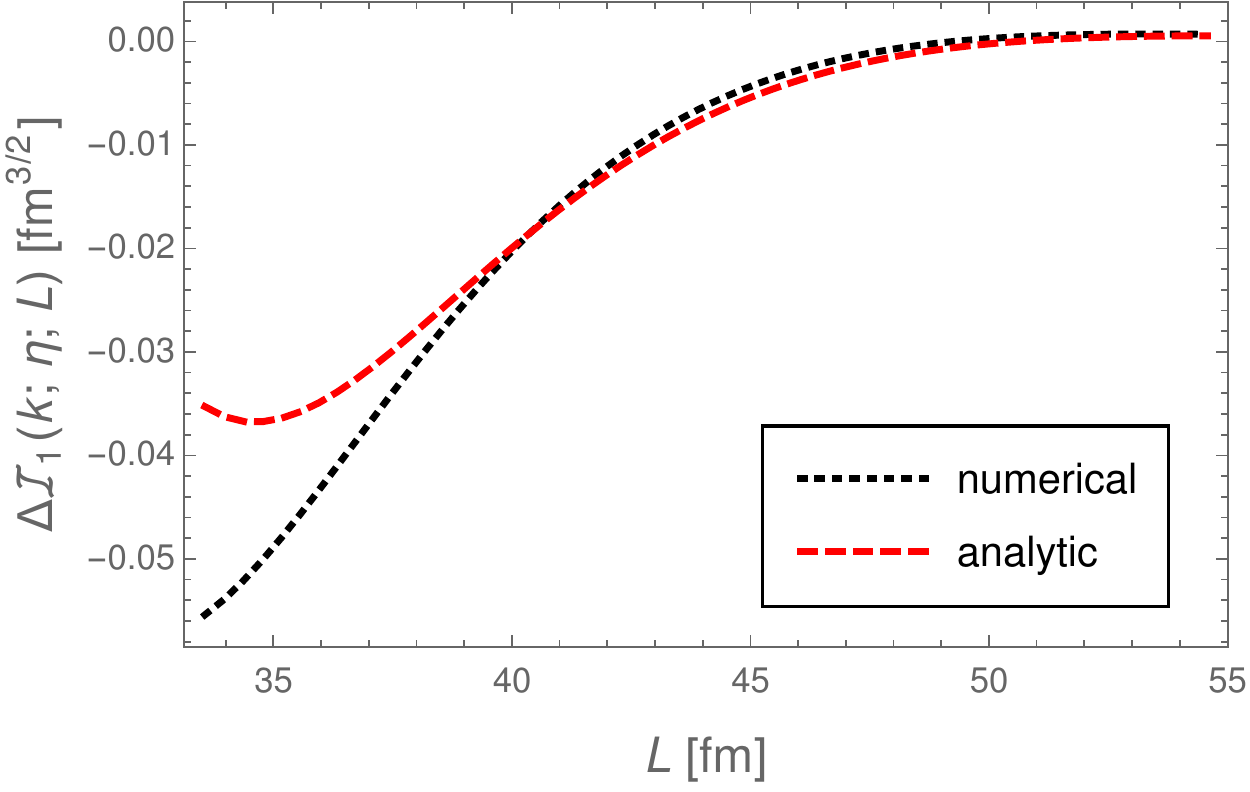}
     \caption{\label{fig:np2M} Numerical and analytic results for the correction
       $\Delta\mathcal{I}_1(k;\eta;L)$ to the radial matrix element of
       the $E1$ operator between the $np$ $^3P_1$ scattering wave
       function at 1~MeV center-of-mass energy and the deuteron
       $s$-wave state.}
\end{figure}

Finally, in Fig.~\ref{fig:np2M}, we plot the IR truncation error in the matrix element of the 
electric dipole ($E1$) operator between the deuteron
$s$-wave and the $np$ $^3P_1$ wave functions, which 
contributes to the radiative $np$ capture,
\begin{equation}
n + p \rightarrow d + \gamma,
\end{equation}
and its reverse process, deuteron photodisintegration. Here the analytic 
formula for $\Delta\mathcal{I}_\lambda(k;\eta;L)$ has neglected terms
from the $\mathcal{O}[1/(kr)]$ corrections to Eq.~\eqref{eq:scatteringwfn}. 
Since these terms are suppressed by a factor of $1/L$, we get a better agreement 
between the analytic and the numerical results at larger $L$ values.

The analytic results shown above in Figs.~\ref{fig:pp100k},
\ref{fig:pp2M} and \ref{fig:np2M} were not fit to the data. The
quantities $A_\infty$, $\gamma_\infty$, and $\delta_l$ were known {\it
  a priori} from the wave functions, and the IR truncation error was
thus completely predicted by Eq.~\eqref{eq:asympt}.  For systems with
$A>2$, however, extracting values for the single-particle separation
energies, asymptotic normalization coefficients and phase shifts might
not be as straight-forward. Moreover, the use of our analytic results
in practical applications is to obtain
$\mathcal{I}_\lambda (k;\eta;\infty)$ by extrapolation when the size
of the basis is constrained due to unavailability of computational
resources.  One computes $\mathcal{I}_\lambda (k;\eta;L)$ at several
large values of $L$, and fits Eq.~\eqref{eq:asympt} [or, if required,
Eq.~\eqref{eq:general}] to these data with
$\mathcal{I}_\lambda (k;\eta;\infty)$, $A_\infty$, $\gamma_\infty$,
and $\delta_l$ treated as fit parameters.  We present the results of
such extrapolations for the $pp$ fusion process in
Table~\ref{tab:fit}. The extrapolations are robust not only at 1~MeV
but also at 50~keV center-of-mass energy, where the neglected
contributions to our extrapolation formula are larger. We found that
the differences in $\mathcal{I}_\lambda (k;\eta;\infty)$ values for
different sets of input data are very small compared to those of the
other fit parameters, $A_\infty$, $\gamma_\infty$, and $\delta_l$
(data not shown). These are not determined very well by the fit
because of the relatively large number of fit parameters in
Eq.~\eqref{eq:asympt}. However, we want to remark that in any standard
calculation these could be fit to several other observables such as
finite volume binding energies or radii thereby increasing the
constraints on these parameters significantly.

 Since Eq.~\eqref{eq:asympt} is valid at asymptotically large values of $L$,
we obtain better fits when the input data contains larger $L$ values.
However, even for smaller $L$, the extrapolation error is much smaller
than the IR truncation error one would make by avoiding extrapolation
and simply using $\mathcal{I}_0 (k;\eta;L_\mathrm{max})$ instead.

\begin{table*}
\centering
\begin{tabular}{cc|ccc|ccc}\hline
 & & \multicolumn{3}{c|}{$\eta=0.50$} & \multicolumn{3}{c}{$\eta=0.11$} \\

$~L_\mathrm{min}$ & $L_\mathrm{max}~$ & $~\mathcal{I}_0 (k;\eta;L_\mathrm{min})~$ & $~\mathcal{I}_0 (k;\eta;L_\mathrm{max})~$ & $~\mathcal{I}_0 (k;\eta;\infty)~$ & $~\mathcal{I}_0 (k;\eta;L_\mathrm{min})~$ & $~\mathcal{I}_0 (k;\eta;L_\mathrm{max})~$ & $~\mathcal{I}_0 (k;\eta;\infty)~$ \\

\hline
15.30 & 20.29 & 0.4063 & 0.4440 & 0.4704 & 2.585 & 2.621 & 2.590 \\
20.29 & 24.28 & 0.4440 & 0.4583 & 0.4712 & 2.621 & 2.611 & 2.592 \\
15.30 & 24.28 & 0.4063 & 0.4583 & 0.4706 & 2.585 & 2.611 & 2.591 \\
15.30 & 39.51 & 0.4063 & 0.4708 & 0.4711 & 2.585 & 2.592 & 2.592 \\
\hline

\hline\hline
  \end{tabular}
\caption{Values of $\mathcal{I}_0 (k;\eta;\infty)$ (in $\mathrm{fm}^{1/2}$) for $pp$ fusion at 50~keV ($\eta=0.50$) and 1~MeV ($\eta=0.11$) center-of-mass energies, 
obtained by fitting Eq.~\eqref{eq:asympt} to the $(L,\mathcal{I}_0 (k;\eta;L))$ data 
for $L$ ranging from $L_\mathrm{min}$ to $L_\mathrm{max}$ (in $\mathrm{fm}$). The fit results agree very well with the 
numerically-approximated values of $\mathcal{I}_0 (k;\eta;\infty)$, which are 0.4711 and 2.592~fm$^{1/2}$ for $\eta=0.50$ and $\eta=0.11$ respectively, for all fit intervals. \label{tab:fit}}
\end{table*}

\paragraph{Summary.}

We studied the dependence of the nucleon capture cross section on the
radius $L$ of the hard wall with Dirichlet boundary condition, which
arises as an effective infrared cutoff when the bound-state wave
function is represented in a truncated basis. We presented an
expression of this radius for computations based on hyperspherical
harmonics.  We showed that the infrared convergence of the cross
section thus calculated is much slower than that of bound state
properties whose errors generally scale as $e^{-2\gamma_\infty
  L}$~\cite{furnstahl2012,more2013,odell2016}.  We also showed that
this feature can lead to errors in the $pp$ fusion cross section that
are comparable in size to uncertainties induced by the nucleon-nucleon
interaction and the electroweak current operator in state-of-the-art
calculations~\cite{marcucci2013,acharya2016}.  We derived a simple
analytic formula for controlled extrapolation of the cross section to
the infinite basis limit. By exploiting our ability to calculate the
two-body wave function for a very wide range of basis size while
concurrently maintaining ultraviolet convergence, we tested our
predictions for two different two-nucleon capture processes. 
Our extrapolation formula also holds for $A>2$ nuclei since their
single particle bound- and scattering-state wave functions also have
the form given respectively by Eqs.~\eqref{eq:boundwfn} and
\eqref{eq:scatteringwfn}. However, for the proton capture process, the
large value of $\eta$ in heavier nuclei restricts the domain of
validity of our extrapolation formula to very high energies. In such case, 
one needs to replace Eq.~(\ref{eq:scatteringwfn}) by the full
Coulomb wave function to compute the IR correction numerically.  An analytic 
derivation of such results, which could facilitate calculations at the energy regime relevant to the
rp-process, is left for future work.

\begin{acknowledgments}

  We thank S. Bacca and L. Marcucci for useful discussions, and S. Binder for comments on the 
  manuscript. This work was supported by the National Science Foundation under
  Grant Nos. PHY-1516077 and PHY-1555030, by the Office of Nuclear Physics,
  U.S.~Department of Energy under Awards Nos.\ DEFG02-96ER40963,
  DE-SC0008499 (NUCLEI SciDAC Collaboration) and under Contract
  No.\ DE-AC05-00OR22725, by the Swedish Research Council under Grant No. 2015-00225, 
  and the Marie  Sklodowska Curie Actions, Cofund, Project INCA 600398. 
  
\end{acknowledgments}


%

\end{document}